\documentclass[a4paper,10pt]{article}

\usepackage[utf8]{inputenc} 
\usepackage{lipsum} 

\usepackage[top=2.5cm, bottom=3cm, left=3.5cm, right=3.5cm,
           heightrounded,marginparwidth=1.5cm, marginparsep=1cm]{geometry} 

\usepackage{changepage} 

\usepackage[shortlabels]{enumitem} 

\usepackage[square,numbers,merge,comma,sort&compress]{natbib} 
\makeatletter
\def\NAT@spacechar{\,}  
\makeatother

\usepackage{amsmath,amssymb,amsfonts,amsthm,amsbsy}
\usepackage{mathtools} 

\usepackage{fnpct}
\setfnpct{dont-mess-around} 

\usepackage{slashed,cancel}
\usepackage{comment}   
\usepackage{relsize}   
\usepackage{setspace}  
\usepackage{moresize}  
\usepackage{epsfig}
\usepackage{latexsym}
\usepackage{mathrsfs,calligra,aurical} 
\usepackage{calc}
\usepackage{float}
\usepackage{appendix}
\usepackage{xargs}
\usepackage{extarrows}
\usepackage{empheq}

\usepackage{tikz}
\usetikzlibrary{scopes,decorations.pathmorphing,patterns,calc,arrows,
                shapes.geometric,shapes.arrows,decorations.markings,plotmarks}
\usepackage{pgfplots}

\usepackage[blocks]{authblk}  

\usepackage[fulladjust]{marginnote}%

\usepackage{graphicx} 
\graphicspath{{Figures/}} 

\usepackage[labelsep=colon]{caption}  
\usepackage[labelsep=colon,aboveskip=10pt,belowskip=10pt]{subcaption}
\usepackage{array,multirow,makecell,booktabs}  
\newcolumntype{M}[2]{>{\centering\arraybackslash$}#1{#2\linewidth}<{$}}
\newcolumntype{T}[2]{>{\centering\arraybackslash}#1{#2\linewidth}<{}}
\newcolumntype{R}[1]{>{\raggedleft\arraybackslash}m{#1\linewidth}<{}}
\newcolumntype{L}[1]{>{\raggedright\arraybackslash}m{#1\linewidth}<{}}
\newcommand\thinrule{\midrule[0.00001pt]}

\makeatletter
\renewcommand\mcell@classz{\@classx
   \@tempcnta \count@
   \prepnext@tok
   \@addtopreamble{
      \ifcase\@chnum
         \hfil
         \mcell@agape{\d@llarbegin\insert@column\d@llarend}\hfil \or
         \hskip1sp
         \mcell@agape{\d@llarbegin\insert@column\d@llarend}\hfil \or
         \hfil\hskip1sp
         \mcell@agape{\d@llarbegin \insert@column\d@llarend}\or
         \mcell@agape{$\vcenter
         \@startpbox{\@nextchar}\insert@column\@endpbox$}\or
         \mcell@agape{\vtop
         \@startpbox{\@nextchar}\insert@column\@endpbox}\or
         \mcell@agape{\vbox
         \@startpbox{\@nextchar}\insert@column\@endpbox}%
      \fi
      \global\let\mcell@left\relax\global\let\mcell@right\relax
    }\prepnext@tok}
\makeatother

\usepackage{titlesec}

\titleformat{\section}{\normalfont\large\bfseries}{\thesection}{0.5em}{}
\titleformat{\subsection}{\normalfont\normalsize\bfseries}{\thesubsection}{0.5em}{}
\titleformat{\subsubsection}{\normalfont\normalsize\bfseries}{\thesubsubsection}{0.5em}{}

\titlespacing*{\section}{0pt}%
                {4ex plus 1ex minus .5ex}{1.75ex plus .25ex minus .25ex}
\titlespacing*{\subsection}{0pt}%
                {3.5ex plus 1ex minus .5ex}{1.25ex plus .2ex minus .2ex}
\titlespacing*{\subsubsection}{0pt}%
                {2.5ex plus 0.75ex minus .2ex}{0.75ex plus .15ex minus .15ex}
\titlespacing*{\paragraph}{0pt}%
                {1.85ex plus 0.5ex minus .15ex}{1em}

\usepackage{titletoc}

\addtocontents{toc}{\addvspace{-0.75em}}  

\titlecontents{section}
  [1.25em] {\addvspace{0.7em plus 0pt}\small}
  {\thecontentslabel\hspace{0.75em}}{}
  {\hspace{0.5em}\titlerule*[0.5em]{.}\contentspage}
  [\addvspace{0.0em plus 0pt}]

\titlecontents{subsection}
  [2.75em] {\addvspace{0.075em plus0pt}\fns}
  {\thecontentslabel\hspace{0.75em}}{\thecontentslabel\hspace{0.75em}}
  {\hspace{0.5em}\titlerule*[0.5em]{.}\small\contentspage}
  [\addvspace{0.075em plus 0pt}]

\usepackage{environ}
\makeatletter
\NewEnviron{subalign}[1]{
\begin{subequations}\label{#1}
\begin{align} \BODY \end{align}
\end{subequations}      }
\makeatother
\makeatletter
\newenvironment{subeqs}%
{\begingroup%
\setlength{\abovedisplayskip}{10pt plus 4pt minus 9pt}%
\setlength{\abovedisplayshortskip}{0pt plus 2pt minus 2pt}%
\setlength{\belowdisplayskip}{12pt plus 3pt minus 9pt}%
\setlength{\belowdisplayshortskip}{7pt plus 3pt minus 4pt}%
\begin{subequations}%
}%
{\end{subequations}\ignorespacesafterend%
\endgroup}%
\makeatother

\makeatletter
\newenvironment{subeqsds}[2]%
{\begingroup%
\setlength{\abovedisplayskip}{{#1}pt plus 2pt minus 9pt}%
\setlength{\abovedisplayshortskip}{0pt plus 0pt minus 2pt}%
\setlength{\belowdisplayskip}{{#2}pt plus 3pt minus 9pt}%
\setlength{\belowdisplayshortskip}{7pt plus 3pt minus 4pt}%
\begin{subequations}%
}%
{\end{subequations}\ignorespacesafterend%
\endgroup}%
\makeatother

\makeatletter
{\begingroup%
\setlength{\abovedisplayskip}{{#1}pt plus 3pt minus 9pt}%
\setlength{\abovedisplayshortskip}{0pt plus 3pt}%
\setlength{\belowdisplayskip}{{#2}pt plus 3pt minus 9pt}%
\setlength{\belowdisplayshortskip}{7pt plus 3pt minus 4pt}%
\begin{equation}%
}%
{\end{equation}\ignorespacesafterend%
\endgroup}%
\makeatother

\makeatletter
{%
\begin{equation}%
\begin{split}%
}%
{\end{split}%
\end{equation}\ignorespacesafterend%
}%
\makeatother

\usepackage{xcolor}
\definecolor{Green}{rgb}{0.05, 0.45, 0.25}
\definecolor{dogwoodrose}{rgb}{0.8, 0.1, 0.55}
\definecolor{RRed}{rgb}{0.7, 0.1, 0.525}

\usepackage{bm}  
\usepackage{dsfont}  

\DeclareMathAlphabet{\mathpzc}{OT1}{pzc}{m}{it}
\DeclareMathAlphabet{\mathcal}{OMS}{cmsy}{m}{n}
\DeclareSymbolFontAlphabet{\Scr}{rsfs}
\DeclareMathAlphabet{\mathbold}{U}{BOONDOX-ds}{m}{n}
\SetMathAlphabet{\mathbold}{bold}{U}{BOONDOX-ds}{b}{n}
\DeclareMathAlphabet{\mathcalboondox}{U}{BOONDOX-calo}{m}{n}
\SetMathAlphabet{\mathcalboondox}{bold}{U}{BOONDOX-calo}{b}{n}
\DeclareMathAlphabet{\mathbcalboondox}{U}{BOONDOX-calo}{b}{n}

\newcommand\linkcol{RRed}



\usepackage[breaklinks=true,backref=page]{hyperref}
\hypersetup{
    bookmarks=true,         
    bookmarksnumbered=true,
    pdftoolbar=true,        
    pdfmenubar=true,        
    pdffitwindow=false,     
    pdfauthor={},     
    pdfsubject={},   
    pdfcreator={},   
    pdfproducer={},  
    pdfpagemode={UseNone},
    pdfstartview={FitH},
    colorlinks=true,
    plainpages,
    linktoc=page,
    citecolor=blue,
    filecolor=black,
    linkcolor=\linkcol,
    urlcolor=Green,
}
\renewcommand*{\backref}[1]{}
\renewcommand*{\backrefalt}[4]{%
\ifcase #1 %
\relax
\or
~{\small [\textsc{p.~\fns{\!#2}}]}
\else
~{\small [\textsc{p.~\fns{\!#2}}]}%
\fi}

\usepackage{footnotebackref}
\usepackage{hypernat} 

\def\+{~+~}
\def\-{~-~}
\def\={~=~}
\newcommand\fns{\footnotesize}

\newcommand\qqquad{\quad\quad\quad}

\newcommand\qLq{\quad\Longrightarrow\quad}
\newcommand\qLrq{\quad\Leftrightarrow\quad}

\newcommand\eps{\varepsilon}
\newcommand\epsz{\varepsilon_\ms{0}}
\newcommand\epsg{\varepsilon_\textrm{g}}
\newcommand\mug{\mu_\textrm{g}}
\newcommand\muz{\mu_\ms{0}}
\newcommand\w{\omega}

\newcommand\hgamma{\bar{\gamma}}

\newcommand\Tc{T_\textrm{c}}
\newcommand\E{\mathbf{E}}
\newcommand\B{\mathbf{B}}
\newcommand\A{\mathbf{A}}
\newcommand\g{\mathbf{g}}
\newcommand\Eg{\mathbf{E}_\textrm{g}}
\newcommand\Bg{\mathbf{B}_\textrm{g}}
\newcommand\Ag{\mathbf{A}_\textrm{g}}
\newcommand\Ee{\mathbf{E}_\textrm{e}}
\newcommand\Be{\mathbf{B}_\textrm{e}}
\newcommand\Ae{\mathbf{A}_\textrm{e}}
\newcommand\jj{\mathbf{j}}
\newcommand\jg{\mathbf{j}_\textrm{g}}
\newcommand\js{\mathbf{j}_\text{s}}

\newcommand\jsk{\mathbf{j}_{\text{s}k}}

\newcommand\kk{\mathbf{k}}
\newcommand\x{\mathbf{x}}

\newcommand\rhog{\rho_\textrm{g}}
\newcommand\phig{\phi_\textrm{g}}

\newcommand\vvs{\mathbf{v}_\text{s}}
\newcommand\vs{v_\text{s}}

\newcommand\cm{\mathrm{cm}}
\newcommand\s{\mathrm{s}}
\newcommand\Kelv{\mathrm{K}}
\newcommand\GN{\mathrm{G}_\ms{\textsc{n}}}
\newcommand\kB{\mathrm{k}_\ms{\textsc{b}}}

\newcommand\Ang{\text{\AA}}

\providecommand{\abs}[1]{\left\lvert#1\right\rvert}

\newcommand{\ms}{\mathsmaller}

\newcommand{\dd}{\partial}
\newcommandx{\tts}[1]{\text{\textsmaller{#1}}}
\newcommandx{\dt}[1][1=f,usedefault]{\frac{\partial{#1}}{\partial t}}
\newcommandx{\dm}[1][1=\mu,usedefault]{\partial_{#1}}
\newcommandx{\dmup}[1][1=\mu,usedefault]{\partial^{#1}}
\newcommandx{\subm}[2][1=p,2=A,usedefault]{{#1}_{\!\mathsmaller{#2}}}
\newcommandx{\subt}[2][1=p,2=A,usedefault]{{#1}_\text{\textsmaller{#2}}}
\newcommandx{\supm}[2][1=p,2=A,usedefault]{{#1}^{\!\mathsmaller{#2}}}
\newcommandx{\supt}[2][1=p,2=A,usedefault]{{#1}^\text{\textsmaller{#2}}}
\newcommandx{\subpt}[3][1=p,2=A,3=B,usedefault]{{#1}^\text{\textsmaller{#3}}_\text{\textsmaller{#2}}}
\newcommandx{\subpm}[3][1=p,2=A,3=B,usedefault]{{#1}^{\mathsmaller{#3}}_{\mathsmaller{#2}}}
\newcommandx{\sh}[1][1=\alpha,usedefault]{\sinh\left(#1\right)}
\newcommandx{\ch}[1][1=\alpha,usedefault]{\cosh\left(#1\right)}
\newcommandx{\sech}[1][1=\alpha,usedefault]{\mathrm{sech}\left(#1\right)}
\newcommandx{\cosech}[1][1=\alpha,usedefault]{\mathrm{cosech}\left(#1\right)} \newcommandx{\LCTd}[4][1=\mu,2=\nu,3=\rho,4=\sigma,usedefault]{\eps_{#1#2#3#4}}
\newcommandx{\LCTu}[4][1=\mu,2=\nu,3=\rho,4=\sigma,usedefault]{\eps^{#1#2#3#4}}

\newcommandx{\gmetr}[2][1=\mu,2=\nu,usedefault]{g_{{#1}{#2}}}
\newcommandx{\invgmetr}[2][1=\mu,2=\nu,usedefault]{g^{{#1}{#2}}}
\newcommandx{\spc}[3][1=\mu,2=a,3=b,usedefault]{{\w_{#1}}^{\!\!{#2}{#3}}}
\newcommandx{\Conn}[3][1=\mu,2=\nu,3=\lambda,usedefault]{{\Gamma_{{#1}{#2}}}^{\!\!#3}}
\newcommandx{\viel}[2][1=\mu,2=a,usedefault]{{e_{#1}}^{\!#2}}
\newcommandx{\inviel}[2][1=a,2=\mu,usedefault]{{e_{#1}}^{#2}}
\newcommandx{\vieluu}[2][1=\mu,2=a,usedefault]{e^{#1#2}}
\newcommandx{\Rdduu}[4][1=\mu,2=\nu,3=a,4=b,usedefault]{{R_{{#1}{#2}}}^{{#3}{#4}}}
\newcommandx{\hgamui}[1][1=0,usedefault]{\hgamma^{\mathsmaller{#1}}}
\newcommandx{\hgamdi}[1][1=0,usedefault]{\hgamma_{{}_{#1}}}
\newcommandx{\gamui}[1][1=0,usedefault]{\gamma^{\mathsmaller{#1}}}
\newcommandx{\gamdi}[1][1=0,usedefault]{\gamma_{{}_{#1}}}

\newcommandx{\emetr}[2][1=\mu,2=\nu,usedefault]{\eta_{{#1}{#2}}}
\newcommandx{\invemetr}[2][1=\mu,2=\nu,usedefault]{\eta^{{#1}{#2}}}
\newcommandx{\hmetr}[2][1=\mu,2=\nu,usedefault]{h_{{#1}{#2}}}
\newcommandx{\invhmetr}[2][1=\mu,2=\nu,usedefault]{h^{{#1}{#2}}}
\newcommandx{\bhmetr}[2][1=\mu,2=\nu,usedefault]{\bar{h}_{{#1}{#2}}}
\newcommandx{\binvhmetr}[2][1=\mu,2=\nu,usedefault]{\bar{h}^{{#1}{#2}}}
\newcommandx{\hud}[2][1=\mu,2=\nu,usedefault]{{h^{#1}}_{\!\!#2}}
\newcommandx{\Ruddd}[4][1=\sigma,2=\mu,3=\lambda,4=\nu,usedefault]{{R^{#1}}_{\!{#2}{#3}{#4}}}
\newcommandx{\Gam}[3][1=\lambda,2=\mu,3=\nu,usedefault]{{\Gamma^{#1}}_{\!{#2}{#3}}}
\newcommandx{\Gamd}[3][1=\mu,2=\nu,3=\lambda,usedefault]{\Gamma_{{#1}{#2}{#3}}}
\newcommandx{\Ricci}[2][1=\mu,2=\nu,usedefault]{R_{{#1}{#2}}}
\newcommandx{\GEinst}[2][1=\mu,2=\nu,usedefault]{G^{{}^\tts{(E)}}_{{#1}{#2}}}
\newcommandx{\Gscr}[3][1=\mu,2=\nu,3=\rho,usedefault]{\mathscr{G}_{{#1}{#2}{#3}}}

\makeatletter
\normalsize
\divide\@tempdima\baselineskip
\@tempcnta=\@tempdima
\makeatother

\DeclareFixedFont\trfont{OT1}{phv}{b}{sc}{11}

\title{%
       \vspace{-1.5cm}
       \centering\boldmath\LARGE\bfseries%
       Exploiting weak field gravity-Maxwell symmetry in superconductive fluctuations regime
       \bigskip
       }

\author{\textsc{Giovanni Alberto Ummarino}
\vspace{0.1em}}
\affil{%
\makebox[\textwidth][c]{Politecnico di Torino, Dipartimento di Scienza Applicata e Tecnologia, corso Duca degli Abruzzi 24, 10129 Torino, Italy}%
}
\affil{National Research Nuclear University MEPhI, Kashirskoe shosse 31, 115409
       Moscow, Russia%
\vspace{-0.025em}
}%
\affil{\href{mailto:giovanni.ummarino@polito.it}{\texttt{giovanni.ummarino@polito.it}}
       }

\smallskip

\author{\textsc{Antonio Gallerati}
\vspace{0.1em}
}
\affil{%
\makebox[\textwidth][c]{Politecnico di Torino, Dipartimento di Scienza Applicata e Tecnologia, corso Duca degli Abruzzi 24, 10129 Torino, Italy}
       }
\affil{Istituto Nazionale di Fisica Nucleare, Sezione di Torino, via Pietro
       Giuria 1, 10125 Torino, Italy%
\vspace{-0.025em}
       }%
\affil{\href{mailto:antonio.gallerati@polito.it}{\texttt{antonio.gallerati@polito.it}}
      }

\date{}

\begin{document}

\maketitle

\smallskip

\begin{abstract}
\noindent
We study the behaviour of a superconductor in a weak static gravitational field for temperatures slightly greater than its transition temperature (fluctuation regime).
Making use of the time-dependent Ginzburg–Landau equations, we find a possible short time alteration of the static gravitational field in the vicinity of the superconductor, providing also a qualitative behaviour in the weak field condition. Finally, we compare the behaviour of various superconducting materials, investigating which parameters could enhance the gravitational field alteration.
\end{abstract}

\bigskip

\tableofcontents



\pagebreak


\section{Introduction} \label{sec:Intro}
It is since 1966, with the paper of DeWitt \cite{DeWitt:1966yi}, that there is great interest in the interplay between the theory of gravitation and superconductivity \cite{Kiefer:2004hv}.
In the following years were produced a lot of theoretical papers about this topic \cite{papini1967detection,Papini:1970cw,rothen1968application,rystephanick1973london,hirakawa1975superconductors,minasyan1976londons,anandan1977gravitational,anandan1979intgra,anandan1984relthe,anandan1994relgra,ross1983london,Felch:1985pre,dinariev1987relativistic,peng1991electrodynamics,peng1990new,peng1991interaction,li1991effects,li1992gravitational,torr1993gravitoelectric,de1992torsion}, until Podkletnov and Nieminem declared to have observed a gravitational shielding in a disk of YBaCuO (YBCO) \cite{podkletnov1992possibility}, an high-$\Tc$ superconductor (HTCS). Of course, after the publication of this paper, other groups tried to repeat the experiment obtaining controversial results
\cite{li1997static,de1995alternative,unnikrishnan1996does,tajmar2009measuring,tajmar2011evaluation,podkletnov2003investigation,poher2017enhanced}, so that the question is still open.\par
Many researchers tried to give a theoretical explanation \cite{ciubotariu1996absence,agop1996gravitational,agop2000some,agop2000local,ivanov1997gravitational,ahmedov1999general,ahmedov2005electromagnetic,Tajmar:2002gm,deMatos:2006tn,Tajmar:2004ww,tajmar2008electrodynamics,ning2004gravitational,chiao2006interface,de2007gravitoelectromagnetism,de2008electromagnetic,de2008gravitational,de2010physical,de2012modified,inan2017interaction,inan2017new,atanasov2017geometric,Sbitnev2019} of the experimental results of Podkletnov and Nieminem in subsequent years, although, in our opinion, the clearest work was made by Modanese in 1996 \cite{modanese1996theoretical,modanese1996role}, interpreting the experimental results in the frame of a quantum field formulation. However, the complexity of the formalism makes it difficult to extract quantitative predictions.\par
In a previous work \cite{Ummarino:2017bvz}, we determined the possible alteration of a static gravitational field in a superconductor making use of the time-dependent Ginzburg–Landau equations \cite{Cyrot_1973,zagrodzinski2003time,alstrom2011magnetic}, providing also an analytic solution in the weak field condition \cite{PhysRevD.39.2825,ruggiero2002gravitomagnetic}. Now, we develop quantitative calculations in a range of temperatures very close but higher than the critical temperature, in the regime of fluctuations \cite{larkin2002fluctuation}.

\section{Weak field approximation} \label{sec: Weak field}
\sloppy
Let us consider a nearly flat spacetime configuration (weak gravitational field) where the metric $\gmetr$ can be expanded as:
\begin{equation}
\gmetr~\simeq~\emetr+\hmetr\;,
\label{eq:gmetr}
\end{equation}
with $\hmetr$ small perturbation of the flat Minkowski metric%
\footnote{%
we work in the mostly plus convention, {$\emetr=\mathrm{diag}(-1,+1,+1,+1)$}}%
. The inverse metric in the linear approximation is given by
\begin{equation}
\invgmetr~\simeq~\invemetr-\invhmetr\;,
\end{equation}
and the Christoffel symbols, to linear order in $\hmetr$ are written as
\begin{equation}\label{eq:Gam}
\Gam[\lambda][\mu][\nu]~\simeq~\frac12\,\invemetr[\lambda][\rho]\,
     \left(\dm[\mu]\hmetr[\nu][\rho]+\dm[\nu]\hmetr[\rho][\mu]-\dm[\rho]\hmetr[\mu][\nu]\right)\;.
\end{equation}
The Riemann tensor is defined as
\begingroup
\setlength{\belowdisplayskip}{8pt plus 2pt minus 2pt}%
\begin{equation}
\Ruddd\= 2\,\dm[{[}\lambda]\Gam[\sigma][\nu{]}][\mu]
        \+2\,\Gam[\sigma][\rho][{[}\lambda]\,\Gam[\rho][\nu{]}][\mu]\;,
\end{equation}
\endgroup
while the Ricci tensor is given by the contraction $\Ricci=\Ruddd[\sigma][\mu][\sigma][\nu]$. To linear order in $\hmetr$, the latter reads \cite{Ummarino:2017bvz}
\begin{equation}
\Ricci~\simeq~\dmup[\rho]\dm[{(}\mu]\hmetr[\nu{)}][\rho]-\frac12\,\dd^2\hmetr-\frac12\,\dm\dm[\nu]h\;,
\label{eq:Ricci}
\end{equation}
with $h=\hud[\sigma][\sigma]$\,.
The Einstein equations \cite{Wald:1984rg,misner1973gravitation} are written as
\begin{equation}
\GEinst\=\Ricci-\dfrac12\,\gmetr\,R
\=8\pi\GN\;T_{\mu\nu}\;,
\end{equation}
and the l.h.s.\ in first-order approximation reads
\begin{equation}
\begin{split}
\GEinst&~\simeq~\dmup[\rho]\dm[{(}\mu]\hmetr[\nu{)}][\rho]-\frac12\,\dd^2\hmetr-\frac12\,\dm\dm[\nu]h
-\frac12\,\emetr\left(\dmup[\rho]\dmup[\sigma]\hmetr[\rho][\sigma]-\dd^2h\right)\;.
\end{split}
\end{equation}
Introducing the symmetric tensor
\begingroup%
\setlength{\abovedisplayshortskip}{2pt plus 3pt}%
\setlength{\abovedisplayskip}{2pt plus 3pt minus 4pt}
\setlength{\belowdisplayshortskip}{5pt plus 3pt}%
\setlength{\belowdisplayskip}{6pt plus 3pt minus 4pt}%
\begin{equation}
\bhmetr\=\hmetr-\frac12\,\emetr\,h\;,
\end{equation}
\endgroup%
the above expression simplifies in \cite{Ummarino:2017bvz}
\begin{equation}
\GEinst~\simeq~
    \dmup[\rho]\dm[{(}\mu]\bhmetr[\nu{)}][\rho]-\frac12\,\dd^2\bhmetr
    -\frac12\,\emetr\,\dmup[\rho]\dmup[\sigma]\bhmetr[\rho][\sigma]
    \=\dmup[\rho]\left(
          \dm[{[}\nu]\bhmetr[\rho{]}][\mu]+\dmup[\sigma]\emetr[\mu][{[}\rho]\,\bhmetr[\nu{]}][\sigma]
              \right)\;.
\end{equation}
If we now define the tensor
\begin{equation} \label{eq:Gscr}
\Gscr~\equiv~
\dm[{[}\nu]\bhmetr[\rho{]}][\mu]+\dmup[\sigma]\emetr[\mu][{[}\rho]\,\bhmetr[\nu{]}][\sigma]\;,
\end{equation}
the Einstein equations can be rewritten in the compact form:
\begin{equation}\label{eq:Einst}
\;\GEinst\=\dmup[\rho]\Gscr\=8\pi\GN\;T_{\mu\nu}\;.
\end{equation}
We can impose a gauge fixing using the \emph{harmonic coordinate condition} \cite{Wald:1984rg}:
\begin{equation}
\Box x^\mu=0
\;\qLrq\;
\dm\left(\sqrt{-g}\,\invgmetr\right)=0
\;\qLrq\;
\invgmetr\,\Gam\,=\,0\;,
\label{eq:gaugefix}
\end{equation}
also called \emph{De Donder gauge}%
\footnote{%
the requirement of a coordinate condition plays the role of a gauge fixing, uniquely determining the physical configuration and removing indeterminacy; in harmonic coordinates, the metric satisfies a manifestly Lorenz-covariant condition, so that the De Donder gauge becomes a natural choice. Moreover, if one considers the weak-field expansion of the EH action in De Donder gauge, the action itself (as well as the graviton propagator) takes a particularly simple form.
}%
. If we now use eqs.\ \eqref{eq:gmetr} and \eqref{eq:Gam} in the last of previous \eqref{eq:gaugefix}, we find, in first-order approximation
\begin{equation}
0~\simeq~ \dm\invhmetr-\frac12\,\dmup[\nu]h\;,
\end{equation}
that is, we have the relations
\begin{equation} \label{eq:gaugecond0}
\dm\invhmetr\simeq\frac12\,\dmup[\nu]h
\;\quad\Leftrightarrow\;\quad
\dmup\hmetr\simeq\frac12\,\dm[\nu]h\;,
\end{equation}
that, in turns, imply the \emph{Lorenz gauge condition}:
\begin{equation}
\dmup\bhmetr~\simeq~0\;.
\end{equation}
The above result simplifies expression \eqref{eq:Gscr} for $\Gscr$, which takes the form
\begingroup%
\setlength{\abovedisplayshortskip}{2pt plus 3pt}%
\begin{equation}
\Gscr~\simeq~\dm[{[}\nu]\bhmetr[\rho{]}][\mu]\;.
\label{eq:Gscr0}
\end{equation}
\endgroup

\subsection{Gravito-Maxwell equations}
Now, let us define the fields \cite{Ummarino:2017bvz}
\begin{subeqsds}{6}{12}\label{eq:fields0}
\begin{align}
\Eg&~\equiv~E_i~=\,-\,\frac12\,\Gscr[0][0][i]~=\,-\,\frac12\,\dm[{[}0]\bhmetr[i{]}][0]\;,\\[\jot]
\Ag&~\equiv~A_i\=\frac14\,\bhmetr[0][i]\;,\\[\jot]
\Bg&~\equiv~B_i
                 \=\frac14\,{\varepsilon_i}^{jk}\,\Gscr[0][j][k]\;,
\end{align}
\end{subeqsds}
where, using \eqref{eq:Gscr0}, we have
\begingroup%
\setlength{\abovedisplayshortskip}{2pt plus 3pt}%
\setlength{\abovedisplayskip}{4pt plus 3pt minus 4pt}
\setlength{\belowdisplayshortskip}{4pt plus 3pt}%
\setlength{\belowdisplayskip}{6pt plus 3pt minus 4pt}%
\begin{equation}
\Gscr[0][i][j]\=\dm[{[}i]\bhmetr[j{]}][0]
\=\frac12\left(\dm[i]\bhmetr[j][0]-\dm[j]\bhmetr[i][0]\right)\=4\,\dm[{[}i]A_{j{]}}\;.
\end{equation}
\endgroup%
First, we find
\begin{equation}
\Bg\=\frac14\,{\varepsilon_i}^{jk}\,4\,\dm[{[}j]A_{k{]}}
   \={\varepsilon_i}^{jk}\,\dm[j]A_k=\nabla\times\Ag
\;\qLq\;
\nabla\cdot\Bg\=0\;.
\end{equation}
Then, one also has
\begin{equation}
\nabla\cdot\Eg\=\dmup[i]E_i~=\,-\dmup[i]\frac{\Gscr[0][0][i]}{2}
              ~=\,-8\pi\GN\;\frac{T_{00}}{2}
              \=4\pi\GN\;\rhog\;,
\end{equation}
using eq.\ \eqref{eq:Einst} and having defined $\rhog\equiv-T_{00}$\,.
If we take the curl of $\Eg$, we obtain
\begin{equation}
\nabla\times\Eg\={\varepsilon_i}^{jk}\,\dm[j]E_k
               ~=\,-{\varepsilon_i}^{jk}\,\dm[j]\frac{\Gscr[0][0][k]}{2}
               ~=\,-\frac14\,4\;\dm[0]\,{\varepsilon_i}^{jk}\,\dm[j]A_k
               ~=\,-\dm[0]B_i~=\,-\frac{\dd\Bg}{\dd t}\;,
\end{equation}
while, for the curl of $\Bg$,
\begin{equation}\label{eq:gravMaxwell4}
\begin{split}
\nabla\times\Bg&\={\varepsilon_i}^{jk}\,\dm[j]B_k
      \=\frac14\,{\varepsilon_i}^{jk}\,
                {\varepsilon_k}^{\ell m}\,\dm[j]\Gscr[0][\ell][m]
      \=\frac12\left(\dmup\Gscr[0][i][\mu]-\dm[0]\Gscr[0][0][i]\right)\=
\\[2\jot]
      &\=\frac12\left(8\pi\GN\;T_{0i}-\dm[0]\Gscr[0][0][i]\right)
      \=4\pi\GN\;j_i+\frac{\dd E_i}{\dd t}
      \=4\pi\GN\;\jg\+\frac{\dd\Eg}{\dd t}\;,
\end{split}
\end{equation}
using again eq.\ \eqref{eq:Einst} and having defined
$\jg \equiv j_i \equiv T_{0i}$\,.\par\smallskip
Following the above prescriptions, we obtained for the fields \eqref{eq:fields0} the set of equations
\begingroup%
\setlength{\abovedisplayshortskip}{2pt plus 3pt}%
\setlength{\abovedisplayskip}{6pt plus 3pt minus 4pt}%
\begin{equation} \label{eq:gravMaxwell}
\begin{split}
\nabla\cdot\Eg&\=4\pi\GN\;\rhog\;;\\[2\jot]
\nabla\cdot\Bg&\=0 \;;\\[2\jot]
\nabla\times\Eg&~=-\dfrac{\dd\Bg}{\dd t} \;;\\[2\jot]
\nabla\times\Bg&\=\frac{4\pi\GN}{c^2}\;\jg
                  \+\frac{1}{c^2}\,\frac{\dd\Eg}{\dd t}\;,\qquad
\end{split}
\end{equation}
\endgroup
having restored physical units \cite{Ummarino:2017bvz}. This equations are formally equivalent to Maxwell equations, with $\Eg$ and $\Bg$ gravitoelectric and gravitomagnetic field respectively.
For example, on the Earth surface, $\Eg$ is simply the Newtonian gravitational acceleration and the $\Bg$ field is related to angular momentum interactions \cite{agop2000local,agop2000some,braginsky1977laboratory,huei1983calculation,peng1990new}.

\subsection{Generalized Maxwell equations}
Now we introduce the generalized electric/magnetic field, scalar and vector potentials, containing both electromagnetic and gravitational terms:
\begin{equation}
\E=\Ee+\frac{m}{e}\,\Eg\,;\quad\;
\B=\Be+\frac{m}{e}\,\Bg\,;\quad\;
\phi=\phi_\textrm{e}+\frac{m}{e}\,\phig\,;\quad\;
\A=\Ae+\frac{m}{e}\,\Ag\,,
\label{eq:genfields}
\end{equation}
where $m$ and $e$ are the mass and electronic charge, respectively, the subscripts identifying the electromagnetic and gravitational contributions.\par
The generalized Maxwell equations for the fields \eqref{eq:genfields} then become \cite{Ummarino:2017bvz,Behera:2017voq}:
\begingroup%
\setlength{\abovedisplayshortskip}{2pt plus 3pt}%
\setlength{\abovedisplayskip}{5pt plus 3pt minus 4pt}%
\begin{equation} \label{eq:genMaxwell}
\begin{split}
\nabla\cdot\E&\=\left(\frac1\epsg+\frac{1}{\epsz}\right)\,\rho \;;\\[2\jot]
\nabla\cdot\B&\=0 \;;\\[2\jot]
\nabla\times\E&~=-\dfrac{\dd\B}{\dd t} \;;\\[2\jot]
\nabla\times\B&\=\left(\mug+\muz\right)\,\jj
                  \+\frac{1}{c^2}\,\dfrac{\dd\E}{\dd t} \;,
\end{split}
\end{equation}
\endgroup%
where $\epsz$ and $\muz$ are the electric permittivity and magnetic permeability in the vacuum, and where have defined
\begingroup%
\setlength{\abovedisplayshortskip}{2pt plus 3pt}%
\setlength{\abovedisplayskip}{5pt plus 3pt minus 4pt}
\begin{equation}
\rhog\=\frac{m}{e}\,\rho\;,\qqquad
\jg\=\frac{m}{e}\;\jj\;,
\end{equation}
\endgroup
$\rho$ and $\jj$ being the electric charge density and electric current density respectively. The new introduced vacuum gravitational permittivity $\epsg$ and vacuum gravitational permeability $\mug$ have the explicit form:
\begingroup%
\setlength{\belowdisplayshortskip}{15pt plus 3pt}%
\setlength{\belowdisplayskip}{20pt plus 3pt minus 4pt}
\begin{equation}
\epsg=\frac{1}{4\pi\GN}\,\frac{e^2}{m^2}\;,\qqquad
\mug=4\pi\GN\,\frac{m^2}{c^2\,e^2}\;.
\end{equation}
\endgroup
We have shown how to define a new set generalized Maxwell equations for
generalized electric $\E$ and magnetic $\B$ fields, in the limit of weak gravitational fields. In the following sections we will use this results to study the behaviour of a superconductor in the fluctuation regime, i.e.\ very close to its critical temperature $\Tc$\,.

\section{The model}
The behaviour of a superconductor in the vicinity of its critical temperature has been extensively studied. This particular region of temperature is characterized by thermodynamic fluctuations of the order parameter, giving rise to a gradual increase of the resistivity of the material from zero to its normal state value, for temperatures $T>\Tc$\,. This happen because, above the critical temperature $\Tc$, the order parameter fluctuations create superfluid regions in which electrons are accelerated. For temperatures larger than $\Tc$, the average size of these regions is much greater than the mean free path, though it decreases with the rise in temperature of the sample.\par
The described regime can be studied by using the time-dependent Ginzburg-Landau equations \cite{Cyrot_1973}. Of course, we have to be sufficiently far from the critical point for this description to be valid (essentially we are dealing with a mean field theory)%
. Moreover, here we suppose we deal with sufficiently dirty materials, in order to observe the effects of the fluctuations over a sizable range of temperature, i.e.\ the electronic mean free path $\ell$ in the normal material has to be less than 10 \Ang.\par
The time-dependent Ginzburg-Landau equations can be written,
for temperatures larger than $\Tc$, with just the linear term, in the gauge-invariant form \cite{hurault1969nonlinear,schmid1969diamagnetic}:
\begin{equation}
\Gamma\left(\hbar\,\dt[]-2\,i\,e\,\phi\right)\psi\=
    \frac{1}{2m}\left(\hbar\,\nabla-2\,i\,e\,\A\right)^2\psi
    +\alpha\,\psi\;,
\end{equation}
where $\psi(\x,t)$ is the order parameter, $\A(\x,t)$ is the potential vector and $\phi(\x,t)$ is the electric potential. Moreover, once defined $\epsilon(T)=\sqrt{\frac{T-\Tc}{\Tc}}$, we also have
\begin{equation}
\alpha=-\frac{\hbar^{2}}{2\,m\,\xi^{2}}\;,\qquad\;
\xi=\xi(T)=\frac{\xi_0}{\sqrt{\epsilon(T)}}\;,\qquad\;
\Gamma=\frac{\abs{\alpha}}{\epsilon(T)}\,\frac{\pi}{8\,\kB\,\Tc}\;,
\end{equation}
\sloppy
where $\xi_0=\xi(0)$ is the BCS coherence lenght.
If we put
\begin{equation}
\psi(\x,t)\=f(\x,t)\,\exp\big(i\,g(\x,t)\big)\;,
\end{equation}
we obtain two equations for the functions $f(\x,t)$ and $g(\x,t)$:
\begingroup%
\setlength{\abovedisplayshortskip}{5pt plus 3pt}%
\setlength{\abovedisplayskip}{10pt plus 3pt minus 4pt}
\setlength{\belowdisplayshortskip}{8pt plus 3pt}%
\setlength{\belowdisplayskip}{12pt plus 3pt minus 4pt}
\begin{subeqs}
\begin{align}
\Gamma\,\hbar\,\dt[f]&\=
    \alpha\,f+\frac{\hbar^{2}}{2m}\,\Delta f-\frac{1}{2}m\,\vs^{2}\,f\;,
\label{eq:dtf}
\\[2.5\jot]
\Gamma\,\hbar\,f\,\dt[g]&\=
    2\,e\,\Gamma\,\phi\,f-\frac{\hbar^{2}}{2m}\,f\,
    \Delta g-2\,\hbar\;\vvs\cdot\nabla f\;,
\label{eq:dtg}
\end{align}
\end{subeqs}
\endgroup
where
\begin{equation}
\vvs=\frac{1}{m}\left(\hbar\,\nabla g+2\,\frac{e}{c}\,\A\right)
\label{eq:vs}
\end{equation}
is the superfluid speed and where the associated current density is
\begin{equation}
\js~=\,-2\,\frac{e}{m}\,|\psi|^{2}\left(h\,\nabla g+2\,\frac{e}{c}\,\A\right)
    ~=\,-2\,e\,f^{2}\,\vvs\;.
\end{equation}
Now, we consider a fluctuation of the wave vector for the function $f$. Let $f_k$ be the value of $f$ for a fluctuation of the wave vector $\kk$. The above equations can be recast in a more convenient form:
\begin{subeqs}
\begin{align}
\Gamma\,\hbar\,\dt[f_{k}]&\=
    \alpha\,f_{k}-\frac{\hbar^{2}}{2m}\,k^{2}\,f_{k}-\frac{1}{2}\,m\,\vs^2\,f_{k}\;,
\label{eq:dtfk}
\\[1.5\jot]
\dt[\vvs]&~=\,-2\,\frac{e}{m}\,\E\;
\label{eq:dtvs}
\end{align}
\end{subeqs}
where the last expression \eqref{eq:dtvs} is obtained by using eq.\ \eqref{eq:vs} and $\nabla\phi=-\E-\frac{1}{c}\dt[\A]$ and taking the gradient of eq.\ \eqref{eq:dtg}.
By integrating \eqref{eq:dtvs} from zero to $t$, we obtain
\begin{equation}
\Gamma\,\hbar\,\dt[f_{k}]\=
    \left(\alpha-\frac{\hbar^{2}}{2m}\,k^{2}-2\,\frac{e^2}{m}\,E^{2}\,t^{2}\right)f_{k}\;,
\end{equation}
so that $f_{k}$ is given by
\begin{equation}
f_{k}(t)\=f_{k}(0)\,\exp\left(\frac{\left(\alpha-\frac{\hbar^2}{2m}k^{2}\right)t-\frac{2}{3}\,\frac{e^2}{m}\,E^{2}\,t^3}{\Gamma\,\hbar}\right)\;,
\end{equation}
with $f_{k}^{2}(0)=\frac{\kB\,T}{2\left(|\alpha|+\frac{\hbar^2}{2m}k^{2}\right)}$ as it was calculated in \cite{de2018superconductivity}. Then, the current $\jsk(t)$ can be written as
\begin{equation}
\jsk(t)\=4\,\frac{e^{2}}{m}\,\E\;t\;f_{k}^{2}(0)\,\exp\left(2\,\frac{\left(\alpha-\frac{\hbar^2}{2m}k^{2}\right)t-\frac{2}{3}\,\frac{e^2}{m}\,E^{2}\,t^3}{\Gamma\,\hbar}\right)\;,
\end{equation}
At this point we sum over $\kk$. The simpler situation is a three-dimensional sample whose dimensions are greater than the correlation length $\xi$, so that we obtain
\begin{equation}
\js(t)\=
    \frac{1}{8\pi^{3}}\,\int_{0}^{+\infty}
       \jsk(k,t)\,4\pi\,k^{2}\,dk\;.
\end{equation}
The potential vector $\A(x,y,z,t)$ can be calculated from:
\begin{equation}
\A(x,y,z,t)=\frac{\mu_0}{4\pi}\int\frac{\js(t)\;\,dx'\,dy'\,dz'}{\sqrt{(x-x')^{2}+(y-y')^{2}+(z-z')^{2}}}\;,
\end{equation}
when the time variations of external fields are small.
The generalized electric field $\E(x,y,z,t)$ of eq.\ \eqref{eq:genfields}, in the case under consideration, can be written as
\begin{equation}
\E(x,y,z,t)~=\,-\dt[\A(x,y,z,t)]+\frac{m}{e}\,\g~=\,-\frac{\mu_0}{4\pi}\dt[\js(t)]\;\mathcal{C}(x,y,z)+\frac{m}{e}\,\g\;,
\end{equation}
where we have considered the static weak (Earth-surface) gravitational field contribution $\g$, and where $\mathcal{C}(x,y,z)$ is a geometrical factor that depends on the shape of the superconductor and on the space point where we calculate the gravitational fluctuations caused by the presence of the superconductor itself. Of course, when $\E=\frac{m}{e}\,\g$ we are in the weak field regime and we can neglect the term proportional to $t^{3}$ in the exponential. In the latter case, for the realisation of an experiment, one needs a weak magnetic field
(we are around $\Tc$) in order to have the superconductor in the normal state, and turn off the magnetic field at the time $t=0$.

\section{Results}
\begin{figure}[H]
\captionsetup{skip=15pt,belowskip=15pt,font=small,labelfont=small,format=hang}
\centering
\includegraphics[width=\textwidth]{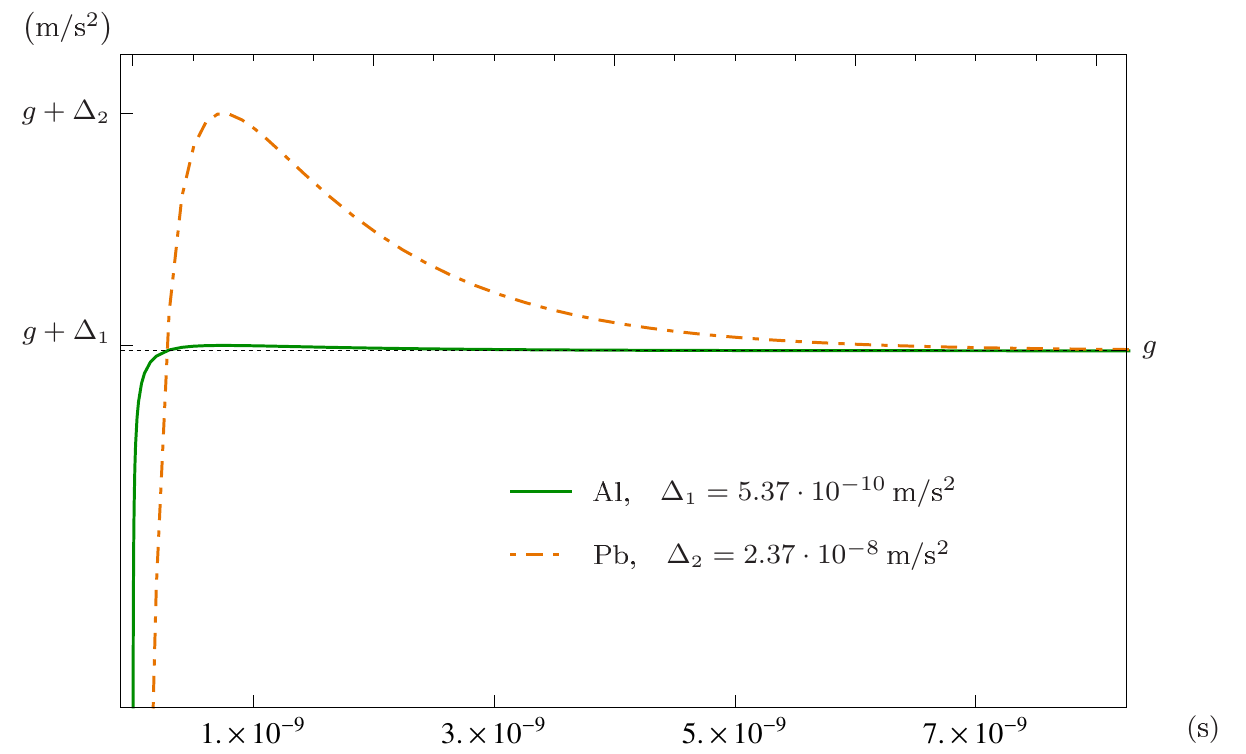}
\caption{The variation of gravitational field as a function of time in the vicinity of a superconductive sample of Al (green solid line) and one of Pb (orange dot-dashed line). The field is measured along the axis of the disk, with bases parallel to the ground, at a fixed distance $d=0.5\,\cm$ above the disk surface. The radius of the disk is $R=10\,\cm$ and the thickness is $h=1\,\cm$.}
\label{fig:APdisk}
\end{figure}

\begin{figure}[H]
\captionsetup{skip=10pt,aboveskip=12pt,belowskip=25pt,font=small,labelfont=small,format=hang}
\centering
\includegraphics[width=\textwidth]{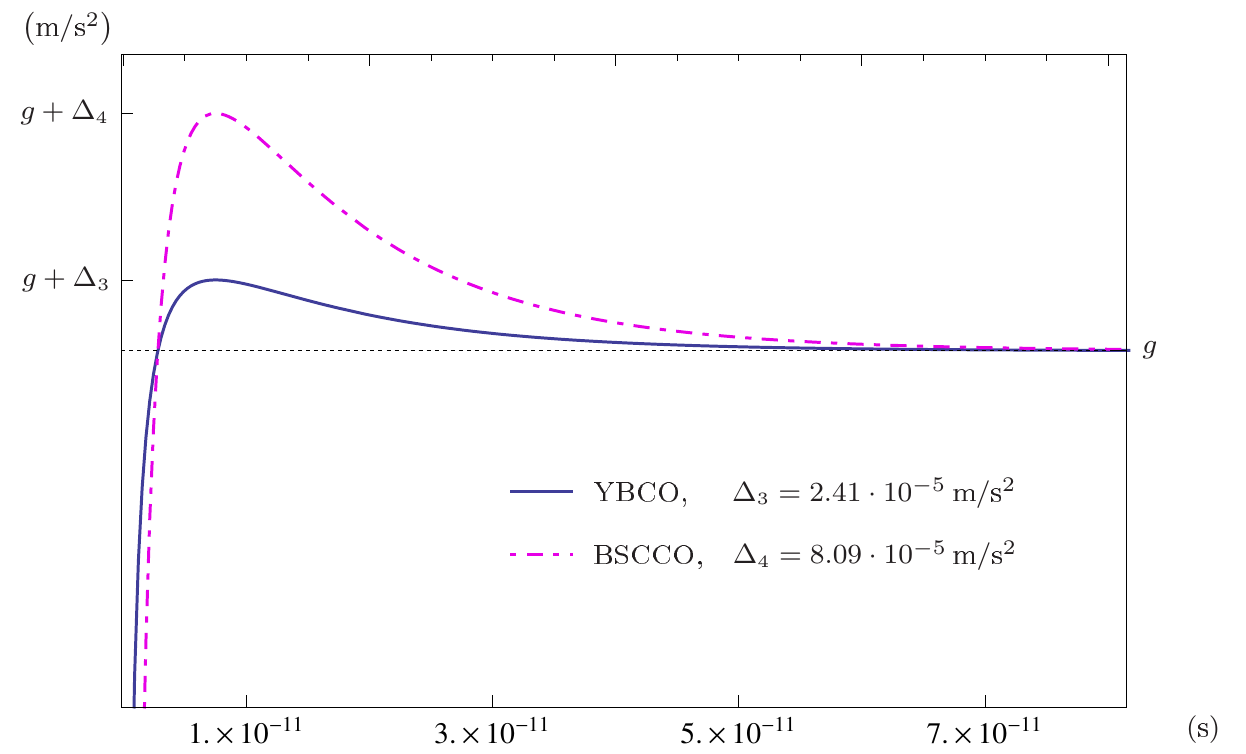}
\caption{The variation of gravitational field as a function of time in the vicinity of a superconductive disk of YBCO (blue solid line) and BSCCO (purple dot-dashed line). The field is measured along the axis of the disk, with bases parallel to the ground, at a fixed distance $d=0.5\,\cm$ above the disk surface. The radius of the disk is $R=10\,\cm$ and the thickness is $h=1\,\cm$.}
\label{fig:YBdisk}
\end{figure}

In Figures \ref{fig:APdisk} and \ref{fig:YBdisk} we show the variation of the gravitational field as a function of time, measured on the axis of a superconductive disk with bases parallel to the ground, at a fixed distance $d$ from the base surface, respectively for low-$\Tc$ (Al and Pb) and high-$\Tc$ superconductors (YBCO and BSCCO). 
The effect is calculated in the range of temperature where superconductive fluctuations are present. The system is initially at a temperature very close to $\Tc$, and we put it in the normal state by using a weak static magnetic field (near $\Tc$ the upper critical field tends to zero). At the time $t=0$, we remove the magnetic field so that the system goes in the superconductive state.\par
It is interesting to note that, in a very short initial time interval, the gravitational field is reduced w.r.t.\ its unperturbed value. After that, it increases up to a maximum value at the time $t=\tau_{0}$ and then decreases to the standard external value. In our previous paper, in the regime under $\Tc$, we found a weak shielding of the external gravitational field \cite{Ummarino:2017bvz}, with the corresponding solution for a simple case. The value $\Delta$ is the maximum variation of the external gravitational field: in principle, field variation is measurable (especially in high-$\Tc$ superconductors), while the problem lies in the very short time intervals in which the effect manifests itself.\par
\begin{figure}[H]
\captionsetup{skip=15pt,belowskip=15pt,font=small,labelfont=small,format=hang}
\centering
\includegraphics[width=\textwidth]{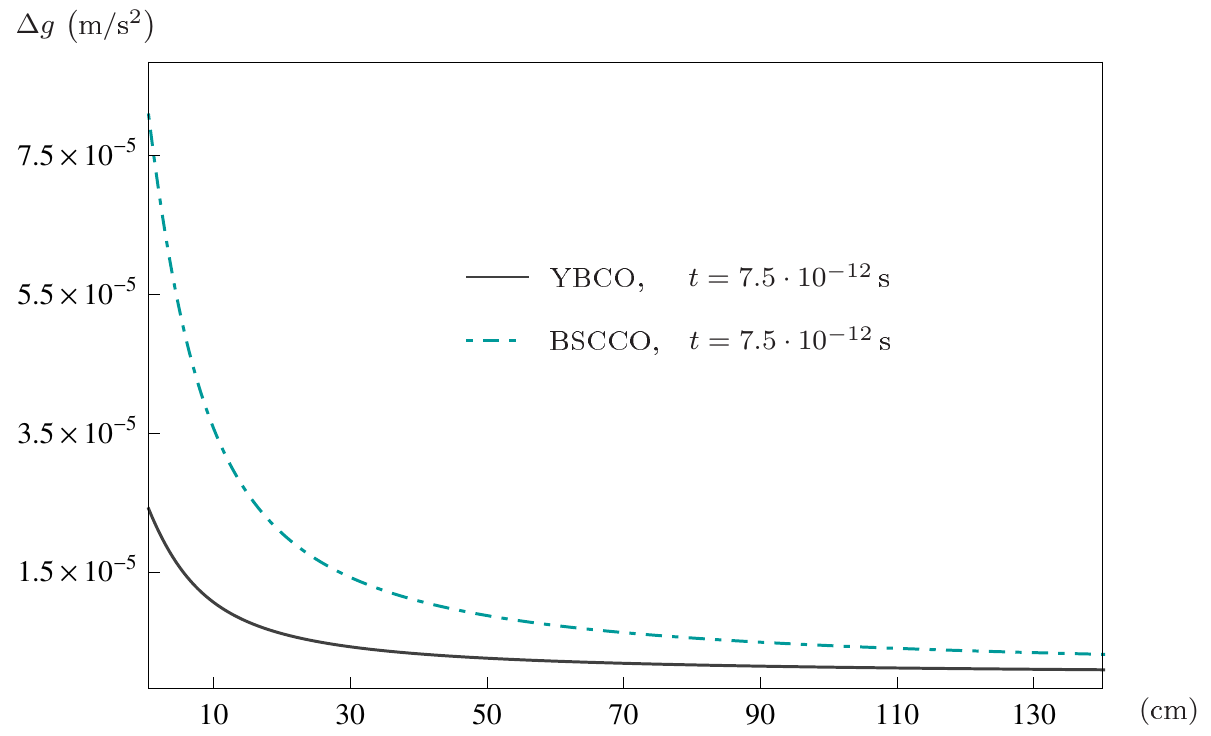}
\caption{The variation of gravitational field as a function of distance in the vicinity of a superconductive sample of YBCO (grey solid line) and one of BSCCO (light blue dot-dashed line). The field is measured along the axis of the disk, with bases parallel to the ground, at the fixed time $t=\tau_0=7.50\cdot10^{-12}\,\s$ that maximizes the variation. The radius of the disk is $R=10\,\cm$ and the thickness is $h=1\,\cm$.}
\label{fig:YBdeltag}
\end{figure}

In Fig.\ \ref{fig:YBdeltag} it is shown the field variation effect as a function of distance from the disk surface, measured along the axis of the disk at the fixed time $t=\tau_0$ that maximizes the effect. In Table \ref{tab:param} we summarize the experimental data for the superconductive materials under consideration.\par
It is instructive to study the values of the parameters that maximize the effect in intensity and time interval. After simple but long calculations, it is possible to demonstrate that $\tau_{0}\propto (T-\Tc)^{-1}$, so it is fundamental to be very close to the critical temperature in order to increase the time range in which the effect takes place. The maximum value of the correction for the external field is obtained for $t=\tau_{0}$ and is proportional to $\xi^{-1}(T)$: this means that the effect is larger in high-$\Tc$ superconductors, having the latter small coherence length.
Clearly this behaviour makes the experimental detection difficult, since if we are close to $\Tc$ we find an increase for the value of $\tau_{0}$ together with a decrease for the alteration of gravitational field, owing to the coherence length divergence at $T=\Tc$\,.
\begingroup
\setlength{\intextsep}{25.0pt plus 2.0pt minus 2.0pt}
\begin{table}[H]
\noindent
\centering
\makegapedcells
\setcellgapes{5pt} 
\begin{tabular}
{@{} T{p}{0.125}| M{p}{0.11} M{p}{0.095} M{p}{0.095} M{p}{0.11} M{p}{0.13}
     M{p}{0.13} }
\toprule
\midrule
     & \Tc\,(\Kelv) & T\,(\Kelv) & \xi_{0}\,\left(\Ang\right) &
       \xi(T)\,\left(\Ang\right) & \tau_{0}\;(\text{s}) & \Delta\,(\text{m}/\text{s}^{2})  \\
\thinrule
  Al    &  1.175  &  1.176  &  15500  &  531313  &  7.45\cdot 10^{-10} &  5.37\cdot10^{-10}
  \\
  Pb    &  7.220  &  7.221  &  870    &  73924   &  7.45\cdot 10^{-10} &  2.37\cdot10^{-8}
  \\
  YBCO  &  89.0   &  89.1   &  30     &  895     &  7.50\cdot 10^{-12} &  2.41\cdot10^{-5}
  \\
  BSCCO &  111.0  &  111.1  &  10     &  333     &  7.50\cdot 10^{-12} &  8.08\cdot10^{-5}
  \\
\midrule
\bottomrule
\end{tabular}
\caption{Input and output parameters for the four different superconductors.}
\label{tab:param}
\end{table}
\endgroup


\section{Conclusions}
We have calculated the possible alteration of a static gravitational field in the vicinity of a superconductor in the regime of fluctuations. We have also shown that the effect should be weak (though perceptible), but it occurs in very short time intervals, making direct measurements difficult to obtain.
Probably some ingredient for a complete depiction of the gravity-superfluid interaction has to be included, as long as it exists, for a more detailed characterization of the phenomenon.\par
Clearly, the goal is to obtain non-negligible experimental evidences (gravitational field perturbations) in workable time scales, trying to optimize contrasting effects by choosing appropriate temperature and sample coherence length. At present, the best option is to choose a HTCS (since very short coherence length increases the intensity of perturbation) and put the system at a temperature very close to $\Tc$ (increase of time range where the effect occurs).
It is also possible that the simultaneous presence of an electromagnetic field with particular characteristics, together with a suitable setting for the geometry of the experiment, could increase the magnitude of the effects under consideration.

\section*{\normalsize Acknowledgments}
\vspace{-5pt}
This work was supported by the MEPhI Academic Excellence Project (contract No.\ 02.a03.21.0005) for the contribution of prof.\ G.\ A.\ Ummarino.
We also thank Fondazione CRT \,\includegraphics[height=\fontcharht\font`\B]{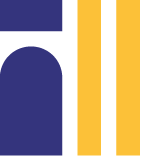}\:
that partially supported this work for dott.\ A.\ Gallerati.


\appendix
\addtocontents{toc}{\protect\addvspace{2.5pt}}%
\numberwithin{equation}{section}%

%


\newpage
\hypersetup{linkcolor=blue}
\phantomsection 
\addtocontents{toc}{\protect\addvspace{3.5pt}}
\addcontentsline{toc}{section}{References} 
\bibliographystyle{mybibstyle}
\bibliography{bibliografia2019ott} 

\end{document}